# Dual-Energy Cone-Beam CT Using Two Complementary Limited-Angle Scans with A Projection-Consistent Diffusion Model

Junbo Peng[1], Chih-Wei Chang[1], Richard L.J. Qiu[1], Tonghe Wang[2], Justin Roper[1], Beth Ghavidel[1], Xiangyang Tang[3] and Xiaofeng Yang[1]*

[1]Department of Radiation Oncology and Winship Cancer Institute, Emory University, Atlanta, GA 30322, USA

[2]Department of Medical Physics, Memorial Sloan Kettering Cancer Center, New York, NY 10065, USA

[3]Department of Radiology and Imaging Sciences and Winship Cancer Institute, Emory University, Atlanta, GA 30322, USA

*Email: xiaofeng.yang@emory.edu

## Abstract

**Background:** Dual-energy imaging on cone-beam CT (CBCT) scanners has great potential in different clinical applications, including image-guided surgery and adaptive proton therapy. However, the clinical practice of dual-energy CBCT (DE-CBCT) has been hindered by the requirement of sophisticated hardware components.

**Purpose:** In this work, we aim to propose a practical solution for single-scan dual-energy imaging on current CBCT scanners without hardware modifications, using two complementary limited-angle scans with a projection-consistent diffusion model.

**Methods:** Our approach has two major components: data acquisition using two complementary limited-angle scans, and dual-energy projections restoration with subsequent FDK reconstruction. Two complementary scans at different kVps are performed in a single rotation by switching the tube voltage at the middle of the source trajectory, acquiring the mixed-spectra projection in a single CBCT scan. Full-sampled dual-energy projections are then restored by a projection-consistent diffusion model in a slice-by-slice manner, followed by the DE-CBCT reconstruction using the FDK algorithm.

**Results:** The proposed method was evaluated in a simulation study of digital abdomen phantoms and a study of real rat data. In the simulation study, the proposed method produced DE-CBCT images at a mean absolute error (MAE) of 20 HU. In the small-animal study, reconstructed DE-CBCT images using the proposed method gave an MAE of 25 HU.

**Conclusion:** This study demonstrates the feasibility of DE-CBCT imaging using two complementary limited-angle scans with a projection-consistent diffusion model in both half-fan and short scans. The proposed method may allow quantitative applications of DE-CBCT and enable DE-CBCT-based adaptive proton therapy.

## 1. Introduction

Dual-energy computed tomography (DECT) exploits the energy-dependent property of X-ray attenuation, resulting in improved capability of material differentiation compared to the conventional single-energy CT (SECT).[1] Since its introduction into medical imaging by Alvarez and Macovski,[2] the DECT has been substantially investigated and become a routine practice in advanced imaging applications, such as automatic bone removal,[3] iodine quantification,[4] virtual monochromatic imaging,[5] and stopping power determination in proton therapy.[6]

Cone-beam CT (CBCT) is a volumetric imaging technique using a flat-panel detector (FPD) and has been widely used to facilitate online image guidance in dentistry,[7] surgical interventions,[8] and radiation therapy.[9] Integrating the dual-energy capability into CBCT may provide significant benefits for various applications, as demonstrated in a number of preliminary studies, including angiographic scanning,[10] material classification,[11] and dose calculation in image-guided photon and proton therapy.[12-14] Despite the increasing research interest in the clinical applications of dual-energy CBCT (DE-CBCT), it has yet to be a routine imaging practice due to the lack of a practical DE-CBCT solution.

In the early stage of DECT implementation, the projection data corresponding to two different X-ray spectra were acquired by carrying out two full CT scans at different tube kVps.[2] With increasing demand for fast data acquisition and rising concerns about patient dose, single-scan DECT is required to minimize the motion-induced spatial misregistration and radiation dose,[15] For DE-CBCT, a single-scan data acquisition is critical because the gantry rotation is slower by almost two orders of magnitude than that of diagnostic multi-detector CT (MDCT).[16] The current single-scan solutions in diagnostic DECT consist of source-side approaches, including dual and fast kVp-switching sources,[17,18] and detector-side approaches, including dual-layer and photon-counting detectors.[19,20] However, these techniques have not been commercialized for DE-CBCT due to the technical challenges in the output power of generic X-ray tube and the reading speed of FPDs.[21] Despite the efforts of DE-CBCT in prototyping the dual-layer FPD, this scheme demands sophisticated hardware upgrades and suffers from its limited spectral separation.[15,22-24]

To implement DE-CBCT on current onboard CBCT scanners without significant hardware modifications, beam-filtration methods have been proposed, which generate dual-energy photons by selectively hardening the X-ray beams via metal foils. For example, a multi-slit beam filter can generate spatially down-sampled projection data in each projection view by filtering the X-ray beams in specific paths, and a customized regularization-based iterative algorithm is employed for image reconstruction from the sparse projection data.[21,25] A rotating filter can generate angularly down-sampled projection by filtering all incident photons at specific view angles, and DE-CBCT images are reconstructed using dual-energy vectorization and bilateral filtering.[26] Either spatial or angular filtration scheme may achieve satisfactory performance in dual-energy imaging and material decomposition. However, both strategies are performed in a full-fan 360° scan, which is inconsistent with the clinical protocols of onboard CBCT scanners in radiation therapy, including short scan and half-fan scan.

In recent years, deep learning has brought new opportunities to the medical imaging community and has shown promising results in many clinical applications, including DECT imaging.[27-29] In a series of works, researchers have investigated dual-energy data acquisition using two complementary limited-angle scans at different kVps,[30-32] which can be easily implemented on current CBCT scanners without modification of hardware components.[29] To reconstruct dual-energy images from angular-limited data acquisition, these works introduce deep learning-based data restoration in the image domain or a hybrid of the image and projection domains. Theoretically, the restoration of data solely in the image domain is suboptimal because of the information lost in the reconstruction process from incomplete projection. The hybrid-domain data restoration is a common way for image formation in CT from incomplete data acquisition, such as limited-angle CT, sparse-view CT, and CT with metal implants.[33] However, this strategy is not suitable for DE-CBCT because it would require excessive GPU memory for large three-dimensional (3D) models and gradient propagation of Feldkamp-David-Kress (FDK) reconstruction layers.[34]

In this work, we propose a framework of projection restoration and image reconstruction for DE-CBCT using two complementary limited-angle scans with a projection-consistency diffusion model. Hard data consistency is enforced into the posterior sampling stage of a diffusion model, enabling accurate data restoration in the projection domain from the acquired mixed-energy data. The dual-energy projections are restored slice by slice, eliminating the large GPU memory requirement by the large 3D models. The feasibility of the proposed method is evaluated and verified via digital phantom study and small animal study, respectively, over the cases of half-fan geometry and short scan.

## 2. Methodology

### 2.1 Overview

Figure 1(a) shows the scanning geometry of the DE-CBCT using two complementary limited-angle scans in short scan and half-fan scan. With the proposed system design, the projections within half of the view angular range are acquired at the high-kVp spectrum and those within the other half angular range are in the low-kVp spectrum.

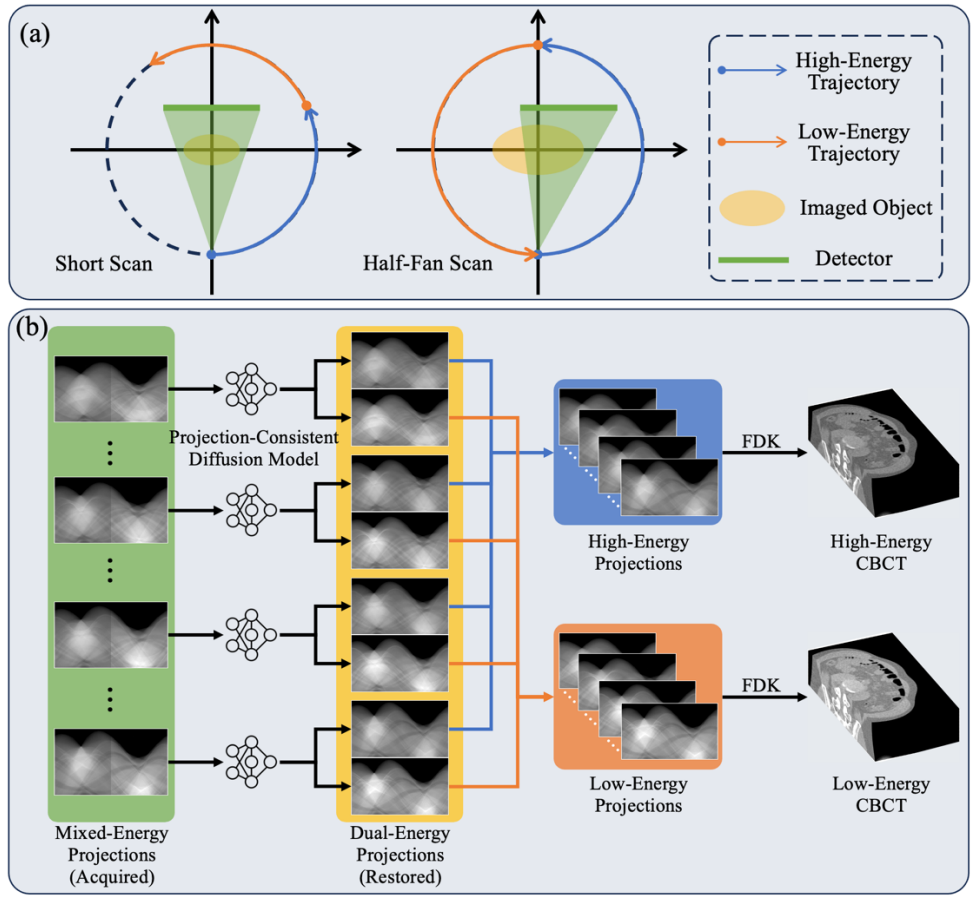

**Figure 1**. The framework of the proposed single-scan DE-CBCT method. (a) Data acquisition using two complementary limited-angle scans. (b) Workflow of the proposed slice-by-slice projection restoration and subsequent DE-CBCT reconstruction using FDK algorithm.

For each detector row, the acquired projection is mixed-energy as shown inside the green block in Figure 1 (b), which is called a projection slice in this work. To restore full-sampled dual-energy projections from the mixed-energy projections within limited GPU memory, the data restoration is performed on each projection slice with a dedicated projection-consistency diffusion model. Figure 1(b) shows the workflow of the proposed projection restoration in a slice-by-slice manner and subsequent DE-CBCT reconstruction using FDK algorithm.

## 2.2 Projection-consistent diffusion model

The diffusion model is an emerging generative approach that achieves state-of-the-art performances in many applications including medical imaging and medical image synthesis.[35-38] In this work, we propose a projection-consistent diffusion model that incorporates data consistency into the dual-energy data restoration in the projection domain.

*2.2.1 Principles of denoising diffusion probabilistic model (DDPM)*

The DDPM is one implementation of the diffusion models that converts a standard Gaussian distribution to the target data distribution via a Markov chain.[39] The DDPM consists of a analytical forward diffusion process and a learning-based reverse denoising process. The forward process starts with a clean image sample in the target distribution, i.e., $x_0 \sim q(x_0)$, and gradually adds Gaussian noise based on the following transition probability

$$q(x_t|x_{t-1}) = \mathcal{N}(x_t; \sqrt{1-\beta_t}x_{t-1}, \beta_t I) \tag{1}$$

where $\mathcal{N}(x_t; \mu, \sigma^2)$ denotes the Gaussian probability distribution function with mean $\mu$ and variance $\sigma^2$, and $\beta_t \in (0,1)$ refers to the scheduled variance for $t = 1, \cdots, T$. By reparameterization of $\alpha_t := 1 - \beta_t$ and $\bar{\alpha}_t := \prod_{i=1}^{t} \alpha_i$, the transition $x_t$ at any time step $t$ can be calculated by

$$x_t = \sqrt{\bar{\alpha}_t}x_0 + \sqrt{1-\bar{\alpha}_t}\epsilon \tag{2}$$

where $\epsilon \sim \mathcal{N}(0, I)$, the noised image $x_T$, becomes nearly an isotropic Gaussian distribution when $T$ is large enough.

Since the posterior of the reverse process $q(x_{t-1}|x_t, x_0)$ depends on the target distribution $q(x_0)$ and is not mathematically tractable, the DDPM tries to learn Gaussian transitions $p_\theta(x_{t-1}|x_t)$ using a neural network with the parameters $\theta$ and $t$ determined as:

$$p_\theta(x_{t-1}|x_t) = \mathcal{N}(x_{t-1}; \mu_{\theta,t}(x_t), \tilde{\beta}_t I) \tag{3}$$

where $\mu_{\theta,t}(x_t)$ refers to the learned mean and $\tilde{\beta}_t := \frac{1-\bar{\alpha}_{t-1}}{1-\bar{\alpha}_t}\beta_t$. Practically, the prediction of $\mu_{\theta,t}(x_t)$ is equivalent to prediction of noise $\epsilon_{\theta,t}(x_t)$, and the posterior in the reverse process can be written as

$$p_\theta(x_{t-1}|x_t) = \mathcal{N}(x_{t-1}; \frac{1}{\sqrt{\alpha_t}}(x_t - \frac{\beta_t}{\sqrt{1-\bar{\alpha}_t}}\epsilon_{\theta,t}(x_t)), \tilde{\beta}_t I) \tag{4}$$

With the learned noise estimator $\epsilon_{\theta,t}(x_t)$ at each time step, one can reverse the diffusion process to recover an image in the target distribution $q(x_0)$ from a white Gaussian noise sample $x_T$ iteratively.

The original DDPM described above is an unconditional generation model, whereas the projection restoration is a generation process conditioned on the acquired mixed-spectra projection. One efficient solution for conditional DDPM is to concatenate the condition image $y$ with the sample $x_t$ along the channel dimension, as such, the learned noise estimator becomes $\epsilon_{\theta,t}(x_t, y^M)$ and the posterior sampling process (Eq. 4) is modified as

$$p_\theta(x_{t-1}|x_t, y^M) = \mathcal{N}(x_{t-1}; \frac{1}{\sqrt{\alpha_t}}(x_t - \frac{\beta_t}{\sqrt{1-\bar{\alpha}_t}}\epsilon_{\theta,t}(x_t, y^M)), \tilde{\beta}_t I) \tag{5}$$

where the condition $y^M$ is the acquired mixed-spectra projection in this context. The forward and reverse diffusion processes are depicted in Figure 2(a).

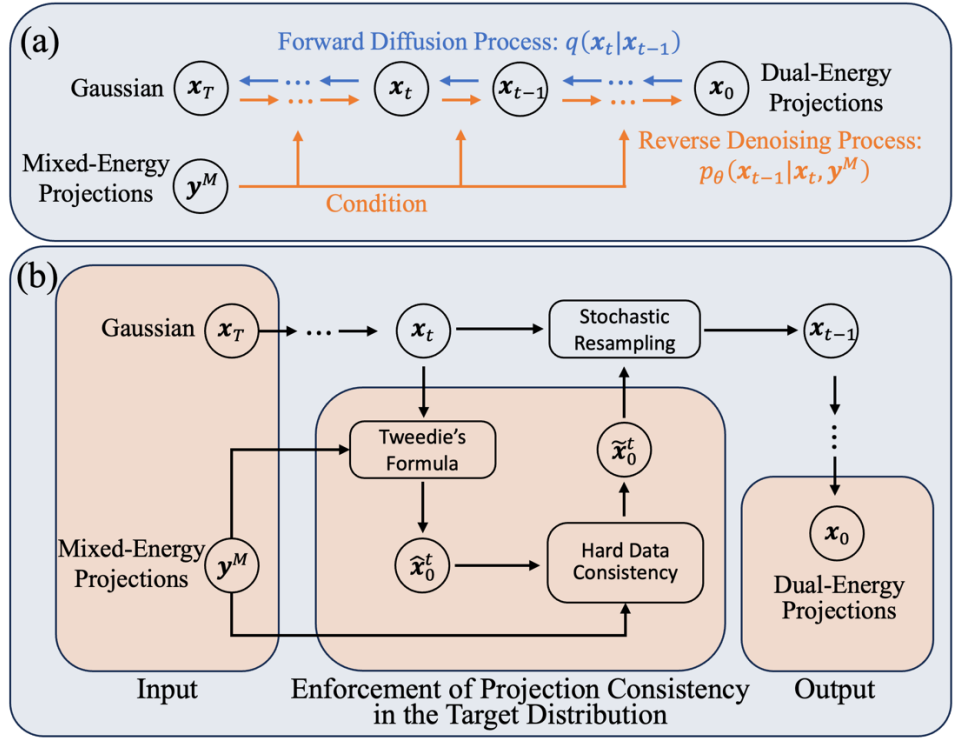

**Figure 2**. Workflow of the proposed projection-consistency diffusion model. (a) The forward and reverse diffusion processes of the conditional DDPM. (b) Dual-energy projections restoration from mixed-energy projection via the projection-consistency diffusion model.

*2.2.2 Hard enforcement of projection data consistency in the target distribution*

Despite the state-of-the-art performance of this conditioning strategy (Eq. 5) in many medical image synthesis applications, the data consistency between $y^M$ and $x_t$ is not enforced during the restoration of dual-energy projections. In other words, the prior information of acquired mixed-spectral projection is not fully utilized in the conditional posterior sampling in Eq. 5.

In the single-scan DE-CBCT using two complementary limited-angle scans, the high- and low-energy parts of the acquired mixed-spectral data should be exactly counterparts in the restored dual-energy projections. Mathematically, such a projection-consistent condition can be formulated as

$$\begin{cases} x_0(H) \otimes \mathcal{M}^H = y^M \otimes \mathcal{M}^H \\ x_0(L) \otimes \mathcal{M}^L = y^M \otimes \mathcal{M}^L \end{cases} \quad (6)$$

where $H/L$ indicates the high- or low-energy channel of the restored dual-energy projections $x_0$, $\mathcal{M}^{H/L}$ refers to the binary mask for each spectrum during the data acquisition, and $\otimes$ is the Hadamard element-wise multiplication operator.

However, such a data fidelity condition only works in the target distribution of $q(x_0)$ instead of all the transition distribution of $q(x_t)$, positioning difficulty to the implementation of Eq. (6) during the reverse denoising process.

To resolve this issue, we first apply Tweedie's formula at each time step,[40] which takes the form

$$\hat{x}_0(x_t) = \frac{1}{\sqrt{\bar{\alpha}_t}} \left( x_t - \sqrt{1 - \bar{\alpha}_t} \cdot \epsilon_{\theta,t}(x_t, y^M) \right) \quad (7)$$

where $\hat{x}_0(x_t)$ denotes the predicted $x_0$ from each transition $x_t$. Then the projection consistency can be enforced at the target distribution of $q(x_0)$ via a proximal optimization problem

$$\tilde{x}_0(x_t) = \text{prox}(\hat{x}_0(x_t))$$
$$= \underset{z}{\arg\min}\|z - \hat{x}_0(x_t)\|_2^2 + \Gamma(z) \tag{8}$$

where $\Gamma(\cdot)$ is an indicator function to implement the hard projection consistency in the form of

$$\Gamma(z) = \begin{cases} 0, \text{if } (z(H) - y^M) \otimes \mathcal{M}^H = 0 \text{ and } (z(L) - y^M) \otimes \mathcal{M}^L = 0 \\ \infty, \text{otherwise} \end{cases} \tag{9}$$

The proximal optimization problem in Eq. (8) can be efficiently solved using the projections-onto-convex sets (POCS) algorithm.[41]

*2.2.3 Stochastic resampling of the noisy transition*

After the enforcement of hard data consistency in the target distribution of $q(x_0)$, the next step is to map the measurement-consistency sample $\tilde{x}_0(x_t)$ back onto the transition data manifold defined by the noisy samples to continue the reverse sampling process.

Recall that the posterior in the reverse process $q(x_{t-1}|x_t, x_0)$ is tractable if both $x_t$ and $x_0$ are available. Based on the Markovian process defined in Eq. (1), we have

$$q(x_{t-1}|x_t, x_0) = \mathcal{N}(x_{t-1}; \tilde{\mu}_t(x_t, x_0), \tilde{\beta}_t I) \tag{10}$$

where

$$\tilde{\mu}_t(x_t, x_0) = \frac{\sqrt{\bar{\alpha}_{t-1}}\beta_t}{1 - \bar{\alpha}_t} x_0 + \frac{\sqrt{\alpha_t}(1 - \bar{\alpha}_{t-1})}{1 - \bar{\alpha}_t} x_t \tag{11}$$

With the optimized projection-consistent $\tilde{x}_0(x_t)$ and the noisy sample $x_t$ at the previous step, we can continue the reverse diffusion process by stochastically resampling the transition $x_{t-1}$ from the parametrized Gaussian distribution

$$p_\theta(x_{t-1}|x_t, \tilde{x}_0(x_t)) = \mathcal{N}(x_{t-1}; \tilde{\mu}_t(x_t, \tilde{x}_0(x_t)), \tilde{\beta}_t I)$$
$$= \mathcal{N}\left(x_{t-1}; \tilde{\mu}_t\left(x_t, \text{prox}\left(\frac{1}{\sqrt{\bar{\alpha}_t}}\left(x_t - \sqrt{1 - \bar{\alpha}_t} \cdot \epsilon_{\theta,t}(x_t, y^M)\right)\right)\right), \tilde{\beta}_t I\right) \tag{12}$$

The proposed strategy for dual-energy projection data restoration with a projection-consistent diffusion model is summarized in Figure 2(b). The pseudo-codes for the conditional training and posterior sampling stages are summarized in Algorithms 1.

**Algorithms 1**. Training and sampling stages of the proposed projection-consistency diffusion model.

| **Algorithm 1.1** Training | **Algorithm 1.2** Sampling | |
|---|---|---|
| 1: **repeat** | 1: $x_T \sim \mathcal{N}(0, I)$ | |
| 2: $(x_0, y) \sim p(x, y)$ | 2: **for** $t = T, \cdots, 1$ **do** | |
| 3: $t \sim U([0,1])$ | 3: $\hat{x}_0(x_t) = \frac{1}{\sqrt{\bar{\alpha}_t}}\left(x_t - \sqrt{1 - \bar{\alpha}_t} \cdot \epsilon_{\theta,t}(x_t, y^M)\right)$ | -Tweedie's formula |
| 4: $\epsilon \sim \mathcal{N}(0, I)$ | 4: $\tilde{x}_0(x_t) = \text{prox}(\hat{x}_0(x_t))$ | -Hard data consistency |
| 5: $x_t = \sqrt{\bar{\alpha}_t}x_0 + \sqrt{1 - \bar{\alpha}_t}\epsilon$ | 5: $x_{t-1} \sim \mathcal{N}(x_{t-1}; \tilde{\mu}_t(x_t, \tilde{x}_0(x_t)), \tilde{\beta}_t I)$ | -Stochastic resampling |
| 6: Take an optimization step on $\nabla_\theta \|\epsilon_{\theta,t}(x_t, y^M) - \epsilon\|^2$ | 6: **end for** | |
| 7: **until** converged | 7: **return** $x_0$ | |

## 3. Evaluation

### 3.1. Image acquisition and preprocessing

Similar to a series of previous works,[29-31] the performance of the proposed DE-CBCT method was evaluated via a simulation study and a small animal study. In the simulation study, digital phantoms were generated from clinical abdomen SECT images, and DE-CBCT scans were simulated in the half-fan geometry. In the small animal study, the DE-CBCT projections were acquired by a flat-panel photon-counting detector and sorted in the view angular range to match the short-scan data acquisition.

*3.1.1 Simulated abdomen scan in half-fan scan mode*

Due to the unavailability of patient DE-CBCT data, the simulated scanning data were generated based on clinical SECT images in this study. The abdomen images were acquired on Siemens SOMATOM Definition AS CT at 120 kVp and reconstructed at $1.0 \times 1.0 \times 1.5$ mm$^3$ voxel size. A total of 4000 abdomen CT slices from 20 patients were used for training and 1000 slices from another 5 patients were used for testing.

There are three major steps in the DE-CBCT data simulation. First, the bone and tissue maps were generated from the abdomen SECT images via segmentation and normalization.[42] Second, the X-ray spectra were simulated with the tube voltages of 140 kVp and 80 kVp at an energy resolution of 1 keV.[43] Third, the dual-energy projections were calculated using Beer-Lambert's law, where Poisson noise was simulated on the detected photons and the unattenuated photon intensity was set at $1 \times 10^6$ per detector pixel.

The scanning was simulated in a half-fan cone-beam geometry with a source-to-detector distance (SID) of 150 cm and a source-to-axis distance (SAD) of 100 cm, in which 360 equiangular projections were acquired over 360 degrees. The FPD had a dimension of 320×200 elements with 1.5×1 mm for each detector element, and the detector is offset to 96 mm from the center.

To acquire the mixed-spectra projection, the measurements from 0 to 180 degrees at 140 kVp were extracted as the high-energy data, and the measurements from 180 to 360 degrees at 80 kVp were taken as the low-energy data.

*3.1.2 Small animal scan in short scan mode*

The DE-CBCT data of ten rats were acquired and used to evaluate and verify the proposed algorithm's performance. The rat scans were performed on a customized micro-CT scanner with a microfocus X-ray tube (L9421-02, Hamamatsu Photonics, Hamamatsu, Japan) and a flat-panel photon-counting detector (XC-Thor, Direct Conversion, Danderyd, Sweden). The X-ray tube was operated at 40 kVp, and the low- and high-energy channels were set at [25.58 33.20) keV and [33.20 40) keV, respectively. The SID and SAD were 222.76 mm and 111.38 mm, and the detector is a 383×64 array with $0.2 \times 0.2$ mm$^2$ element size. Projections were acquired at 600 equiangular positions at a 0.6-degree interval. A 3D volume with 384×384×32 voxels was reconstructed at voxel size $0.1 \times 0.1 \times 0.1$ mm$^3$.

For the short-scan mode, the projections within 216° angular range out of the 360° range were used, which is equivalent to 360 projection views out of the 600 projection views. The projection data of eight rats were used for model training, and the other two rat scans were used for testing. For each rat scan in the training set, 360 projections at 1 to 360, 61 to 420, 121 to 480, and 241 to 600 views were sorted out as the short-scan data. In this way, the training data set was augmented by 4 times. In the testing data set, the acquired data at 151 to 510 projection views were extracted to be consistent with a short scan.

## 3.2. Metrics for image quality assessment

The proposed method was compared with two other DDPM-based methods in the simulation and small animal studies. For quantitative evaluations, we calculated the mean absolute error (MAE), peak signal-to-noise ratio (PSNR), and normalized correlation coefficient (NCC) between the results of different methods and the ground-truth DE-CBCT images. The metrics are defined as

$$MAE = \frac{1}{n_x n_y} \sum_{i,j}^{n_x,n_y} |CT(i,j) - CT_{GT}(i,j)| \qquad (13)$$

$$PSNR = 10 \times \log_{10}\left(\frac{MAX^2}{\frac{1}{n_x n_y}\sum_{i,j}^{n_x,n_y}|CT(i,j) - CT_{GT}(i,j)|^2}\right) \qquad (14)$$

$$NCC = \frac{1}{n_x n_y} \sum_{i,j}^{n_x,n_y} \frac{(CT(i,j) - \overline{CT})(CT_{GT}(i,j) - \overline{CT_{GT}})}{\sigma_{CT}\sigma_{CT_{GT}}} \qquad (15)$$

where $CT(i,j)$ and $CT_{GT}(i,j)$ are the value of voxel $(i,j)$ in the reconstructed and ground-truth CT images, respectively, and $n_x n_y$ is the total number of voxels. $MAX$ is the maximum pixel value in the CT images. $\overline{CT}$ and $\overline{CT_{GT}}$ are the means of reconstructed and ground-truth CT images. $\sigma_{CT}$ and $\sigma_{CT_{GT}}$ are the standard deviations of reconstructed CT and ground-truth images. MAE is the magnitude of the voxel-based Hounsfield unit (HU) difference between the reconstructed and the ground-truth CT images. The PSNR measures if the reconstructed CT intensity is evenly or sparsely distributed, while the NCC is a measure of similarity between the reconstructed and the ground truth CT images as a function of displacement.

### 3.3. Comparison studies with other DDPM-based algorithms

To demonstrate the superiority of the proposed projection-consistency diffusion model for dual-energy projection restoration, we compared it with two other DDPM-based methods.

The first one is to train an unconditional DDPM (Eq. 4) and perform diffusion posterior sampling via the proposed projection-consistent (PC) strategy (2.2.2 and 2.2.3). The second one is to train a DDPM conditioned on the mixed-spectra projection $y^M$ (Eq. 5), but without the data consistency during the reverse sampling stage.

### 3.3. Implementation details

A single model with time embedding,[39] $\epsilon_{\theta,t}(\cdot)$, was trained for all $T$ steps. A U-net shape structure with attention modules and residual blocks was used as the noise estimator.[44] For all three DDPMs, the total number of time steps, $T$, was set to 1000 and the noise variance was linearly scheduled from $1\times10^{-4}$ to $2\times10^{-2}$. The models were trained using the Adam optimizer with a learning rate of $1\times10^{-4}$, betas of 0.9 and 0.999, and eps of $1\times10^{-8}$. The batch size was fixed at 2 and the drop-out ratio was set to 0.3 during the training.

All the CBCT images were reconstructed by the FDK algorithm with a Hamming. All the experiments were conducted using PyTorch 2.1.0 on an 80GB Nvidia A100 GPU. The training was stopped after $2\times10^6$ iterations, which took about 70 h for the simulation study and 85 h for the small animal study. For the sampling stage, it took about 1 minute per projection slice in the simulation study and 1 minute per projection slice in the small-animal study.

## 4. Results

### 4.1. Simulation study

Selected slices of projection data are shown in Figure 3. For each slice, the left column is for the high-energy spectrum and the right one is for the low-energy spectrum. The top row lists acquired data in the single DE-CBCT scan, where the projection is angular-limited in each spectrum. The second to fourth rows summarize the restored dual-energy projections using different diffusion models. The ground-truth data are shown in the bottom row for reference. There is no visible difference between the proposed results and the ground truth. For the dual-energy projections generated from conditional DDPM, residual noises are observed, and the pixel values are biased.

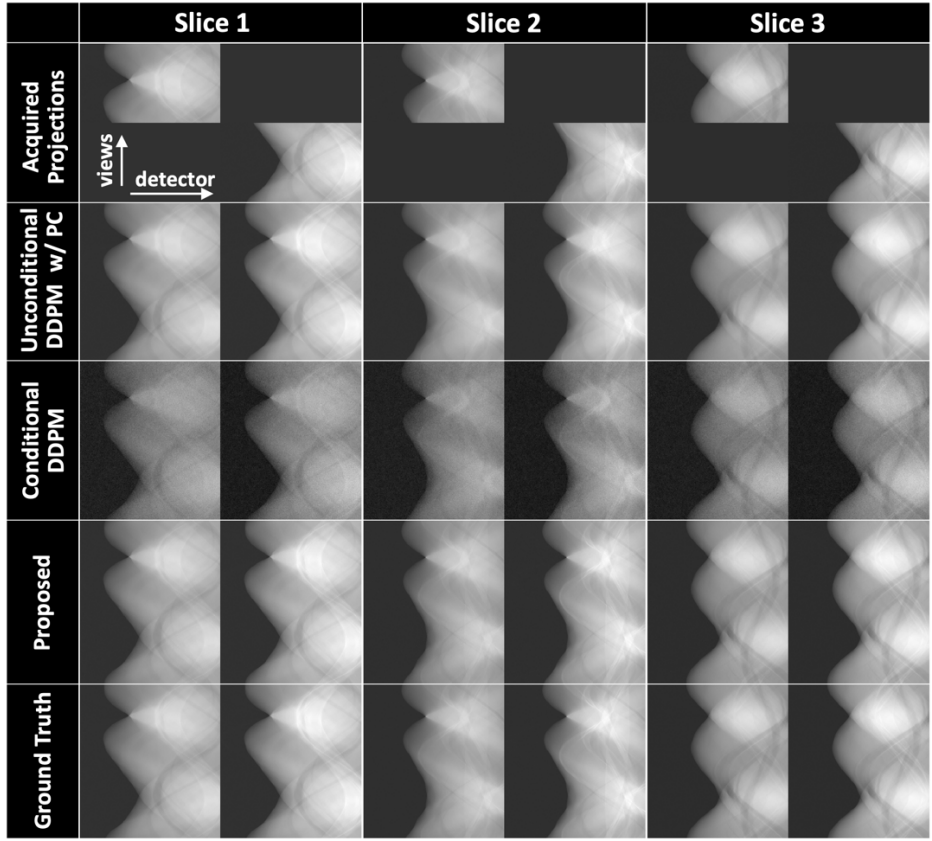

**Figure 3**. Three slices of acquired, restored, and ground-truth projections in the digital abdomen phantom study. For each slice, the left column is for the high-energy data and the right one is for the low-energy data. The top row lists the acquired projections using two complementary limited-angle scans. The second row lists the restored dual-energy projections using unconditional DDPM with the projection-consistent (PC) strategy. The third and fourth rows are the results of conditional DDPM and the proposed projection-consistency diffusion model. The display window is adaptive for each projection slice.

Figure 4 summarizes DE-CBCT images of the digital abdomen phantoms reconstructed from acquired and restored projections using different DDPMs. The first column shows the FDK results from limited-angle data. All three DDPM-based methods improve the image quality from the limited-angle results, while the proposed method provides the best performance of DE-CBCT imaging. The results of unconditional DDPM with PC show blurring anatomical structures and some distortions in the bone-tissue boundaries. The results of conditional DDPM are significantly corrupted by the noise, and anatomical information is nearly totally lost in the low-contrast tissues. The proposed method produces the best results which cannot be visibly differentiated from the ground-truth images.

Table I tabulates the values of quantitative metrics of DE-CBCT images reconstructed using different methods. Consistent with the visual quality results, unconditional DDPM with PC, conditional DDPM, and the proposed method can all improve the quantitative results from the limited-angle FDK images. Among these, the unconditional DDPM with PC outperforms the conditional DDPM in all three metrics, while the proposed method further improves the MAE and PSNR compared to the unconditional DDPM with PC.

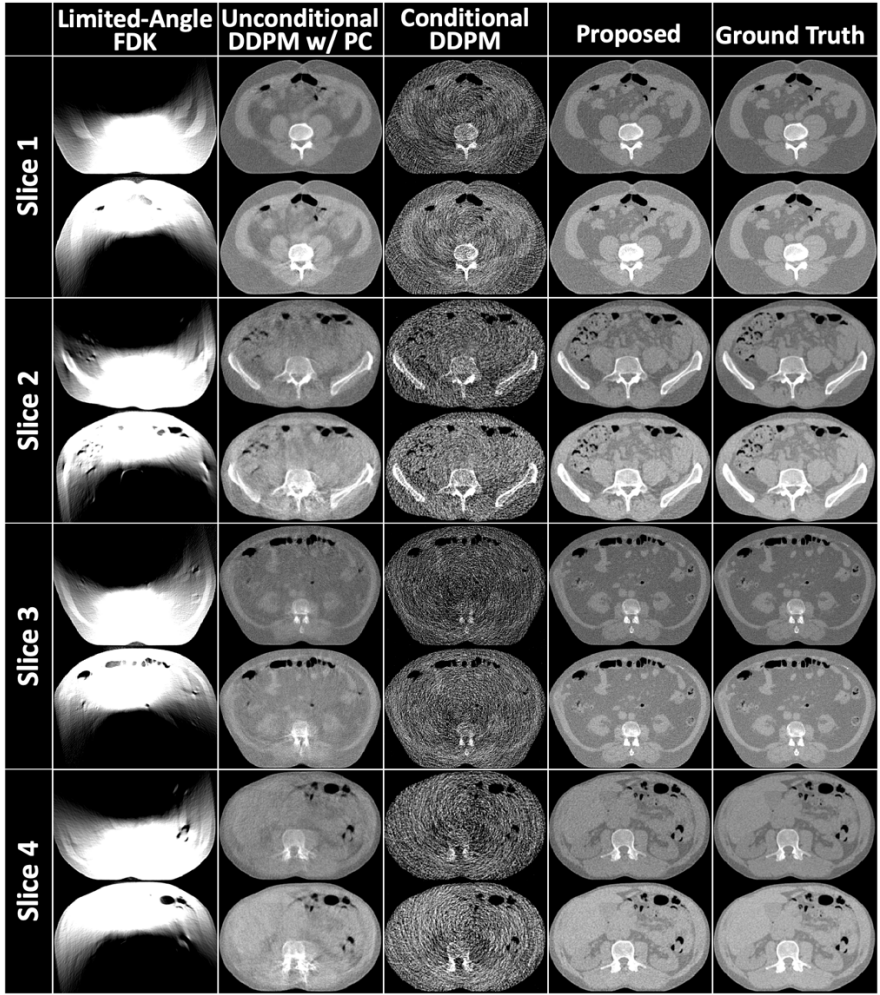

**Figure 4.** Reconstructed DE-CBCT images of the digital abdomen phantoms using different methods. For each slice, the upper row is for the high-energy data and the bottom row is for the low-energy data. The display window is [-500 500] HU.

**Table I**. Numerical comparison among the ground-truth DE-CBCT and reconstructed images using different methods in the simulation study. Bold text indicates the best value in each metric. HE/LE denotes the high- or low-energy.

|  |  | Limited-Angle FDK | Unconditional DDPM w/ PC | Conditional DDPM | Proposed |
|---|---|---|---|---|---|
| **MAE (HU)** | HE-CBCT | 407.58±89.76 | 36.48±3.48 | 159.90±26.60 | **20.36±0.80** |
|  | LE-CBCT | 459.34±102.46 | 42.25±3.73 | 172.25±29.69 | **20.88±0.88** |
| **PSNR (dB)** | HE-CBCT | 12.21±1.38 | 34.70±0.87 | 22.33±1.33 | **40.07±0.32** |
|  | LE-CBCT | 11.10±1.38 | 32.92±0.85 | 21.70±1.38 | **39.75±0.35** |
| **NCC** | HE-CBCT | 0.58±0.05 | 0.99±0.01 | 0.91±0.03 | **1.00±0.00** |
|  | LE-CBCT | 0.49±0.06 | 0.99±0.01 | 0.93±0.02 | **1.00±0.00** |

## 4.2. Small animal study

The performances of projection restoration using different methods in the real rat study are summarized in Figure 5. For each projection slice, the left column is for the high-energy spectrum and the right one is for the low-energy spectrum. As in the simulation study, acquired limited-angle projections are listed in the top row. The second to fourth rows list the restored projections using different methods, and the ground truth is shown in the bottom row for reference. Blurring is observed in the results of unconditional DDPM with projection-consistency sampling. Alteration of the pixel value and spectral inconsistency is observed in the results of conditional DDPM. For the proposed method, there is no visible difference between the results and the ground truth.

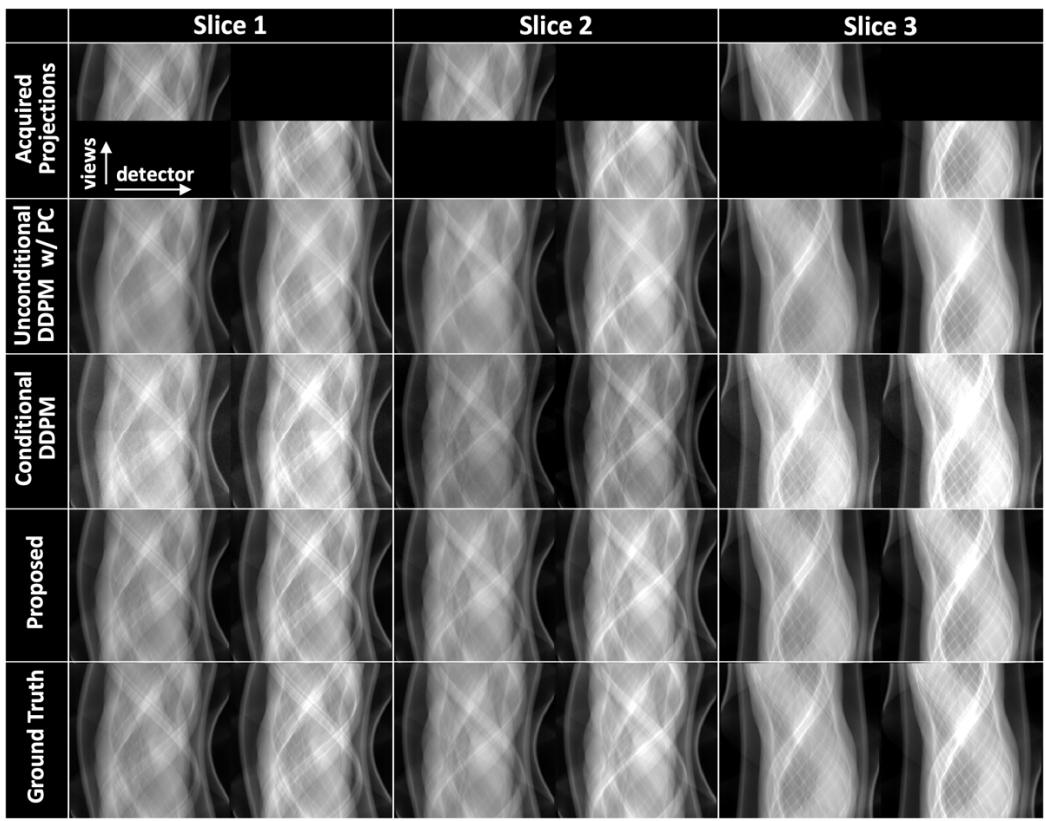

**Figure 5**. Selected slices of acquired, restored, and ground-truth projections in the small-animal study. For each slice, the left column is for the high-energy data and the right one is for the low-energy data. The top row lists the acquired projections using two complementary limited-angle scans. The second row lists the restored dual-energy projections using unconditional DDPM with PC. The third and fourth rows are the results of conditional DDPM and the proposed projection-consistency diffusion model. The display window is adaptive for each projection slice.

Figure 6 shows selected slices of reconstructed real rat DE-CBCT images using different methods. Consistent with the results of projection restoration, artifacts and blurring are observed in the results of unconditional DDPM with PC, and significant HU errors are observed in the results of conditional DDPM. The results of the proposed method show the best HU accuracy and fine-structure preservation, verifying its feasibility in clinical practice.

The results of the quantitative analysis are summarized in Table II, which are consistent with the visual quality comparison. For the high-energy CBCT, the proposed method decreased MAE from 186.33 HU in the limited-angle results to 24.51 HU, increased PSNR from 18.79 dB to 35.77 dB, and improved NCC from 0.76 to 0.99. For the low-energy CBCT, MAE was decreased from 221.46 HU to 26.05 HU, PSNR and NCC were improved from 17.05 dB and 0.75 to 34.75 dB and 1.00 by the proposed method.

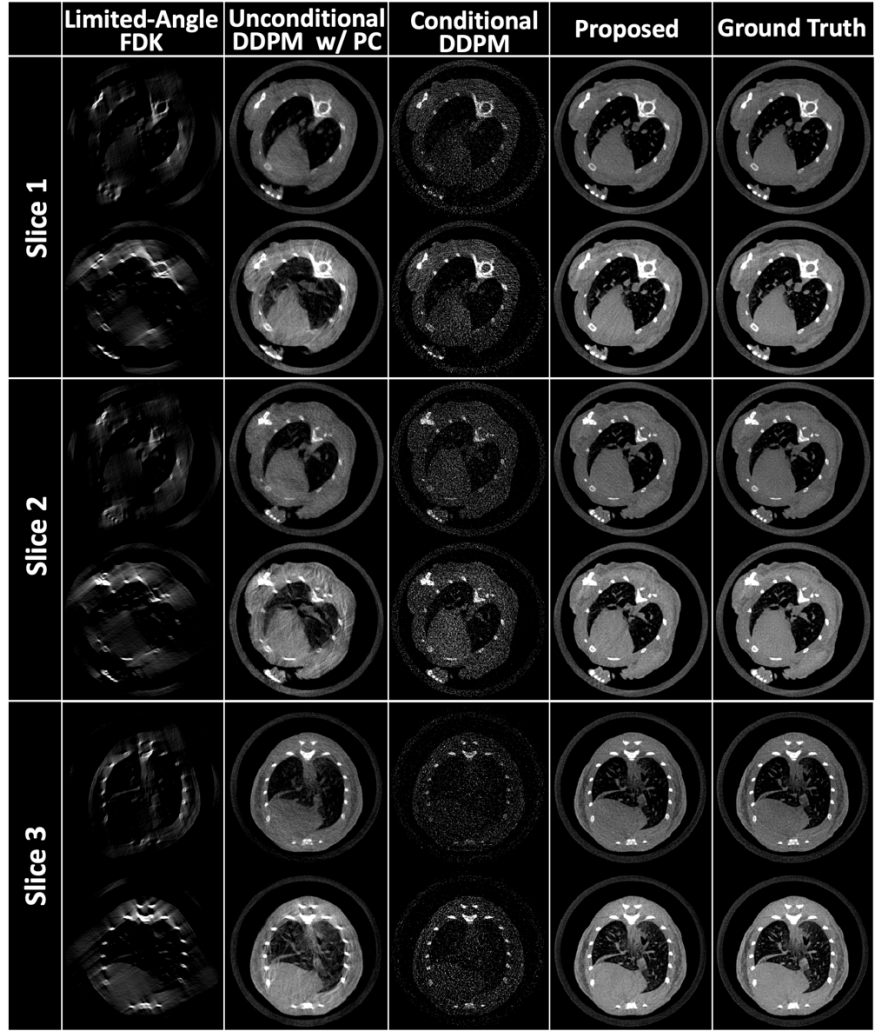

**Figure 6**. Reconstructed DE-CBCT images of the real rats using different methods. For each slice, the upper row is for the high-energy data and the bottom row is for the low-energy data. The display window is [-500 500] HU.

**Table II**. Quantitative comparisons of the reconstructed DE-CBCT using different strategies. Bold text indicates the best value in each metric. HE/LE denotes the high- or low-energy.

|  |  | Limited-Angle FDK | Unconditional DDPM w/ PC | Conditional DDPM | Proposed |
|---|---|---|---|---|---|
| **MAE (HU)** | HE-CBCT | 186.33±8.76 | 48.69±2.85 | 199.53±101.60 | **24.51±1.12** |
|  | LE-CBCT | 221.46±12.29 | 62.67±5.07 | 224.71±117.42 | **26.05±1.89** |
| **PSNR (dB)** | HE-CBCT | 18.79±0.44 | 29.41±0.60 | 18.60±4.20 | **35.77±0.42** |
|  | LE-CBCT | 17.05±0.34 | 26.38±0.56 | 17.66±4.31 | **34.75±0.45** |
| **NCC** | HE-CBCT | 0.76±0.02 | 0.97±0.01 | 0.68±0.17 | **0.99±0.00** |
|  | LE-CBCT | 0.75±0.02 | 0.96±0.01 | 0.72±0.16 | **1.00±0.00** |

## 5. Discussion

This work proposed a practical solution for single-scan dual-energy imaging on current CBCT scanners using two complementary limited-angle scans, with an efficient data restoration method in the projection domain using a projection-consistent diffusion model. With the restored full-sampled dual-energy projections, accurate DE-CBCT images can be reconstructed using analytical algorithms.

There are two primary contributions presented in this work. First, it is the first study to demonstrate the feasibility of DE-CBCT using two complementary limited-angle scans in both half-fan scan and short scan protocols, paving the way to the clinical practice of dual-energy imaging on current onboard CBCT scanners without hardware modifications. Second, we proposed a projection-consistent diffusion for the data restoration of dual-energy projections in a slice-by-slice manner. Such a strategy enables accurate DE-CBCT reconstruction and tackles the challenge imposed by the large GPU memory requirement for cone-beam projection restoration.

The conditional DDPM has shown state-of-the-art performance in a variety of medical imaging applications.[36,37,45] However, as shown in this work, its performance is inferior to the unconditional DDPM with PC in both studies. This means that the proposed projection-consistent strategy is key to the success of the proposed DE-CBCT method, which is within anticipation because the prior information of the mixed-energy projection is fully leveraged in the posterior sampling stage via the hard enforcement of data consistency.

Compared to the previously reported image-domain data restoration scheme for DE-CBCT using two complementary limited-angle scans,[29] there are two advantages associated with the proposed projection-domain data restoration strategy. First, complete anatomical information is preserved in the acquired mixed-spectra projection, but inevitable information loss occurs in the corrupted CT images if they are directly reconstructed from the mixed-spectra projection. Second, full-sampled dual-energy projections with identical ray paths can be generated in the proposed scheme, making it applicable for more projection-domain material decomposition algorithms, e.g., the one-step and two-step decomposition methods.[2,46] It is also worth noting that some SECT-based material decomposition and DECT estimation methods were recently reported.[27,47] However, these methods would require 3D models and projection operators in the cone-beam scan, demanding significantly higher GPU memory for their applications in CBCT imaging.

The contamination of Compton scattering is one of the primary challenges in CBCT, resulting in altered projection data and artifacts in reconstructed images. Similar to previous studies in DE-CBCT,[21,25,29] the Compton scattering is excluded in the simulation study because the objective of this study is focused on image formation, rather than scatter correction, in DE-CBCT. There are two potential solutions to incorporate scatter correction into the proposed DE-CBCT framework. The first one is to remove the scattering in the projection domain before the data restoration, which can be model-based or deep learning-based. The other one is to perform synthetic CT generation based on the reconstructed DE-CBCT using GAN or diffusion models. The incorporation of scatter correction into the DE-CBCT will be the focus of our follow-up studies.

A common issue of diffusion models is the time-consuming sampling procedure, which impedes the adoption of diffusion model-based methods in clinical practice. How to accelerate the sampling of diffusion models has become a hot topic in deep learning research, and previous research has evaluated the performance of different high-efficiency diffusion models in medical imaging tasks, e.g., low-dose CT and PET denoising.[48,49] We will further investigate how to incorporate the acceleration strategy into the proposed projection-consistency diffusion model.

## 6. Conclusions

In this work, we developed a framework of projection restoration and image reconstruction for single-scan DE-CBCT using two complementary limited-angle scans with a projection-consistent diffusion model. The proposed method enables dual-energy imaging on current onboard CBCT scanners without hardware modification and at the radiation dose and acquisition time comparable to that of a single-energy scan. Accurate DE-CBCT images can be formed using the proposed method, allowing advanced quantitative and adaptive applications of DE-CBCT in image-guided clinical procedures, such as surgical interventions and radiation therapy, and particularly, DE-CBCT-based adaptive proton therapy.

## Acknowledgment

The authors extend their sincere appreciation to Professors Shouhua Luo and Yikun Zhang at Southeast University for generously providing the small-animal image data.